\documentclass[pra,aps,superscriptaddress,amsmath,amssymb,twocolumn,reprint]{revtex4-2}
\usepackage{CJKutf8}
\usepackage{graphicx}
\usepackage{dcolumn}
\usepackage{bm,braket}
\usepackage{mathptmx}
\usepackage{ulem}
\usepackage{amsmath}
\usepackage{amsfonts}
\usepackage{amssymb}
\usepackage{mathrsfs}
\usepackage{xcolor}
\usepackage[colorlinks=true,linkcolor=blue,anchorcolor=red,citecolor=blue, urlcolor=blue]{hyperref}
\usepackage{ulem}
\usepackage{pifont}

\makeatletter
\renewcommand*\env@matrix[1][\arraystretch]{%
  \edef\arraystretch{#1}%
  \hskip -\arraycolsep
  \let\@ifnextchar\new@ifnextchar
  \array{*\c@MaxMatrixCols c}}
\makeatother

\def\be{\begin{equation}}
\def\ee{\end{equation}}
\def\ba{\begin{eqnarray}}
\def\ea{\end{eqnarray}}

\newlength{\seplinewidth}
\newlength{\seplinesep}
\colorlet{sepline}{orange}

\begin{document}
\begin{CJK*}{UTF8}{gbsn}
	
	\title{Logarithmic Quantum Time Crystal}
	
	\author{Haipeng Xue (薛海鹏)}
	\thanks{These two authors contributed equally.}
	\affiliation{International Center for Quantum Materials, School of Physics, Peking University, Beijing 100871, China}

	\author{Lingchii Kong (孔令琦)}
	\thanks{These two authors contributed equally.}
	\affiliation{International Center for Quantum Materials, School of Physics, Peking University, Beijing 100871, China}
	
	\author{Biao Wu (吴飙)}%
	\email{wubiao@pku.edu.cn}
	\affiliation{International Center for Quantum Materials, School of Physics, Peking University, Beijing 100871, China}
	\affiliation{
		Wilczek Quantum Center, School of Physics and Astronomy, Shanghai Jiao Tong University, Shanghai 200240, China}
	\affiliation{
		Collaborative Innovation Center of Quantum Matter, Beijing 100871, China}%

	\date{\today}
	
	\begin{abstract}
		We investigate a time-independent  many-boson system, whose ground states are
		quasi-degenerate and become infinitely degenerate in the thermodynamic limit.  Out of these quasi-degenerate ground states
		we construct a  quantum state that  evolves in time  with a period that is logarithmically proportional
		to the number of particles, that is, $T\sim \log N$.  This boson system in such a  state is  a quantum time crystal
		as it approaches the ground state in the thermodynamic limit. The logarithmic dependence of its  period on the total particle number $N$
		makes it observable experimentally even for systems with very large number of particles.
		Possible experimental proposals are discussed.
	\end{abstract}
	
	\maketitle
	
\end{CJK*}

\section{\label{sec:1}INTRODUCTION }
Spontaneous symmetry breaking is a well known physical phenomenon, where the observed ground state of a
many-particle system does not possess the symmetries of its Hamiltonian\cite{beekman2019introduction}, e.g.,
ferromagnetic materials break the rotational symmetry and crystals break the spatial translational symmetry.
Wilczek suggested the possibility of  spontaneous breaking of time translational symmetry\cite{shapere2012classical,wilczek2012quantum}:
the observed ground state of a closed quantum system may oscillate periodically in time. This suggestion
had drawn some quick criticism\cite{nozieres2013time,bruno2013impossibility}.
In 2015, a no-go theorem was proved to exclude the possibility of spontaneous continuous time
translational symmetry breaking in the ground state for a wide class of
Hamiltonians with short-range interactions\cite{watanabe2015absence}. However, it was realized later
that quantum time crystals can exist in a periodically driven system\cite{sacha2015modeling}. This is now
known as discrete time crystal; its equilibrium state can vary with a period that is multiple of the driving period\cite{else2016floquet,khemani2016phase,yao2017discrete}.
Two experiments with trapped ions and nitrogen-vacancy centers were performed, confirming the existence of discrete time crystals\cite{choi2017observation,zhang2017observation}.
Recently,   the discussion about time crystals has expanded to the systems with long-range interaction\cite{kozin2019quantum}.

When spontaneous symmetry breaking occurs in a time independent system,  the observed ground state
is not the true ground state but a superposition of many (quasi-)degenerate ground states\cite{zhu2015extended}.
Two energy states are quasi-degenerate
if the energy gap between them approaches zero in the thermodynamic limit.
There have some efforts to construct a time crystal with these (quasi-)degenerate ground states\cite{shapere2019regularizations,huang2018symmetry,syrwid2021can}.
For such a time crystal,  it typically oscillates with a period that grows polynomially with particle number, i.e. $T\sim O(N^{n})$.
It will be impossible in experiments to keep a many-particle state from decoherence for such a long time.

In this work we study a time-independent many-boson system. This boson system has many
quasi-degenerate ground states, which become infinitely degenerate in the thermodynamic limit.
With these ground states, we are able to construct states that oscillate with time with a period $T$ that
is logarithmically proportional to the number of particles $N$, that is, $T\sim \log N$.
Due to this logarithmic dependence, the period is short enough for  possible experimental observation even for very large $N$.
We find that the observables that are usually used in time crystal experiments are not suitable for our systems.
We suggest to observe the time crystal by measuring the square of overlap
between the states of such a quantum time crystal at different times, i.e., $\left|\langle \Psi(0)|\Psi(t)\rangle\right|^2$ with a technique
based on the Hong–Ou–Mandel interference\cite{hong1987measurement}.


\section{\label{sec:2} Two-mode interacting boson systems}
The interacting system of $N$ bosons that we are going to consider has only two modes. In a certain parameter range, this system
has many quasi-ground states, which approach the true ground state in the thermodynamics limit $N\rightarrow\infty$.
This is the critical feature of this system and a necessary condition for spontaneous symmetry breaking to occur.

\subsection{Theoretical model}\label{sec:2A}
The two-mode interacting boson system is described by the following
Hamiltonian\cite{liang2009atom,cao2012quantum,zhu2015extended}
\begin{align}\label{eq:1}
	\hat{H} &= -(\hat{a}^{\dagger}_{1}\hat{a}_{2}+\hat{a}^{\dagger}_{2}\hat{a}_{1})+
	\frac{\gamma}{N}[\hat{n}_{1}(\hat{n}_{1}-1)+\hat{n}_{2}(\hat{n}_{2}-1)]\notag\\
	&\quad+\frac{4\gamma}{N}\hat{n}_{1}\hat{n}_{2}+
	\frac{\gamma}{N}(\hat{a}^{\dagger}_{1}\hat{a}^{\dagger}_{1}\hat{a}_{2}\hat{a}_{2}+
	\hat{a}^{\dagger}_{2}\hat{a}^{\dagger}_{2}\hat{a}_{1}\hat{a}_{1})\,,
\end{align}
where $\hat{a}^{\dagger}_{1}(\hat{a}_{1})$ and $\hat{a}^{\dagger}_{2}(\hat{a}_{2})$
are the creation (annihilation) operators of two modes and $N$ is the total  number of bosons.
This simple theoretical model can now be realized in experiments\cite{HemmerichBosons}.
In the above, we have set the strength of single particle hopping as the unit of energy and therefore $\gamma$
is a dimensionless parameter characterizing the interaction strength and the pair hopping.
In addition, for simplicity,  we set $\hbar\equiv 1$.

\begin{figure}[h]
	\centering
	\includegraphics[width=1\linewidth]{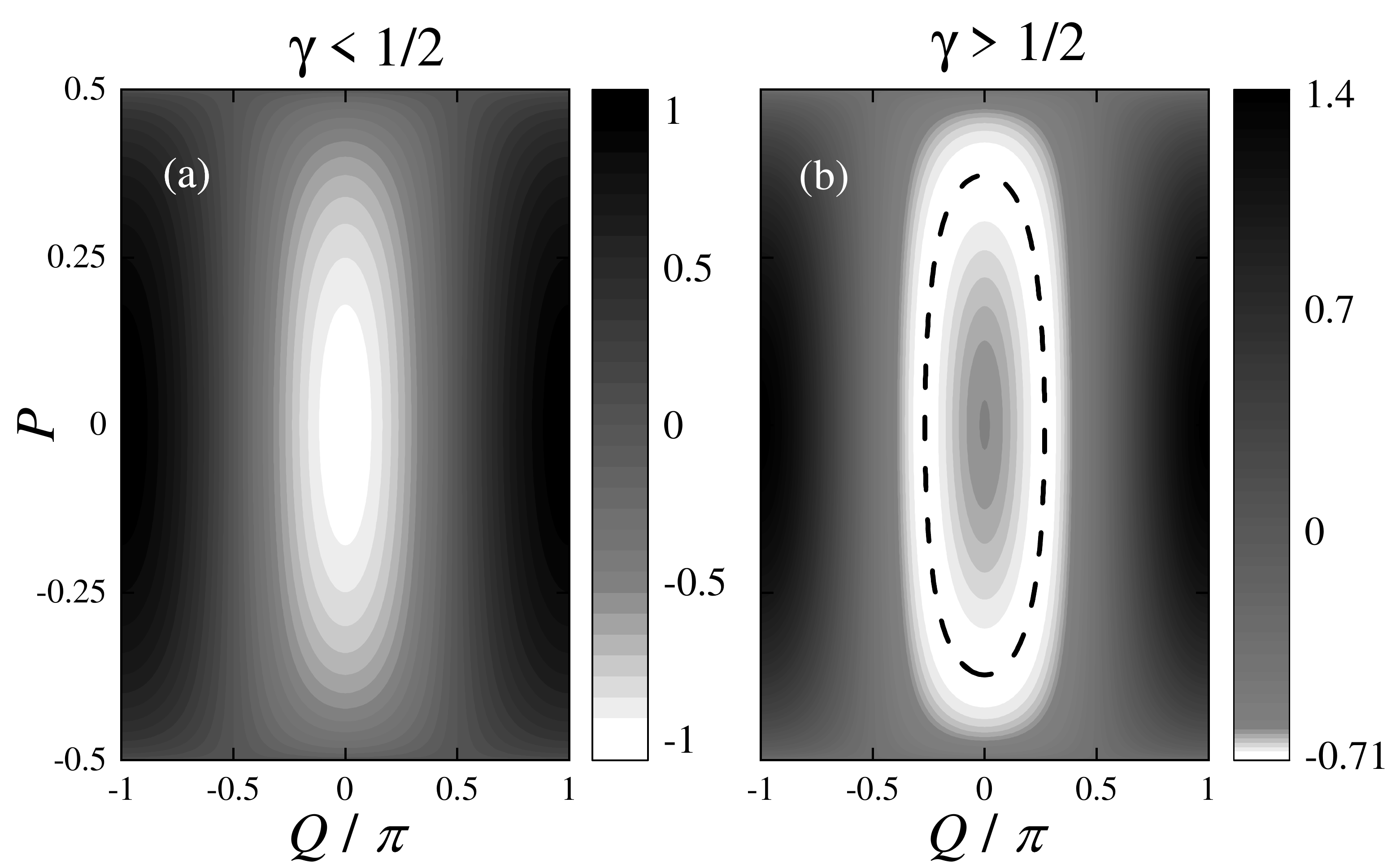}
	\caption{The energy landscape  in the phase space of the mean-field model $H_m$.
	(a) When $\gamma<1/2$, the system has only one  ground state at $(Q,P)=(0,0)$; (b) when $\gamma>1/2$,
	the system has infinitely degenerate ground states, which are located on the dashed line.}
	\label{fig:1}
\end{figure}

In the thermodynamic limit $N\rightarrow\infty$, this quantum many-body system can be described by a mean-field model.
Mathematically, one can simply replace the operators $\hat{a}^\dagger_{1,2}$ and $\hat{a}_{1,2}$ with their corresponding
complex variables $a^*_{1,2}$ and $a_{1,2}$ in Eq.\;\eqref{eq:1}\cite{zhu2015extended}. We choose a different set of variables
$P=(|a_{2}|^2-|a_{1}|^2)/2$ and $Q=\arg(a_{2})-\arg(a_{1})$.  Physically, $P$ and $Q$ are
the particle number difference and the phase difference between the two modes, respectively.
In terms of $P$ and $Q$,  the mean-field model of the system is
\be
\label{eq:2}
H_m = -\sqrt{1-4 P^2}\cos Q +\gamma \left(1-4 P^2\right)\cos^{2}Q\,.
\ee
Note that $P$ and $Q$ are a pair of dynamical variables that are canonically conjugate to each other.
Fig.\;\ref{fig:1} is the energy landscape in the phase space of this mean-field Hamiltonian. When $\gamma <1/2$, the center
of the phase space $P=Q=0$ has the lowest energy.  When $\gamma>1/2$,  all the points on the dashed line
in Fig.\;\ref{fig:1}(b)  have the same lowest energy. That means that the system is  infinitely degenerate
when $\gamma>1/2$. As we shall see in the next subsection, this infinite degeneracy corresponds to
a set of quasi-degenerate ground states in the quantum model \eqref{eq:1}.

\subsection{Quantum energy levels}
\label{sec:2B}
For mathematical simplicity,  we transform this boson system
into a mathematically equivalent spin system.  With $\hat{n}_{1,2}=\hat{a}^{\dagger}_{1,2}\hat{a}_{1,2}$,
we introduce $\hat{S}_{x}=(\hat{a}^{\dagger}_1\hat{a}_2+\hat{a}^{\dagger}_2\hat{a}_1)/2,\quad \hat{S}_{y}=(\hat{a}^{\dagger}_1\hat{a}_2-\hat{a}^{\dagger}_2\hat{a}_1)/2i,\quad \hat{S}_{z}=(\hat{n}_1-\hat{n}_2)/2$. The quantum model \eqref{eq:1} becomes
\begin{equation}\label{eq:3}
\hat{H}^\prime=\hat{H}/2=-\hat{S}_{x}+\frac{2\gamma}{N}\hat{S}^{2}_{x}\,.
\end{equation}
This Hamiltonian is also called Lipkin-Meshkov-Glick model and describes a spin system with $S=N/2$ \cite{lipkin1965validity}.
In this spin formalism the energy eigenstates and energy levels are rather obvious.
As $\hat{S}_{x}$ commutes with $\hat{H}$, the eigenstates of the system are the eigenstates of $\hat{S}_{x}$,
\be
\hat{S}_{x}|m\rangle=m|m\rangle\,,~~m=-S,-S+1,\cdots,0,\cdots,S\,.
\ee
For eigenstate $\ket{m}$,  its corresponding energy level  is
\begin{equation}\label{eq:4}
	E_{m}=\frac{2\gamma}{N}m^2-m\,.
\end{equation}
When $\gamma <1/2$, the lowest energy level is at $m=S=N/2$ and the energy level increases monotonically as $m$ decreases.
And the energy gap $\delta E_m$ between two neighboring levels $\ket{m-1}$ and $\ket{m}$ is
\be
\delta E_m=1-\frac{2\gamma}{N}(2m-1)> 1-2\gamma\,.
\ee
This means that the energy gap remains finite even in the thermodynamic limit $N\rightarrow \infty$.

\begin{figure}[h]
	\centering			
	\includegraphics[width=3.3in]{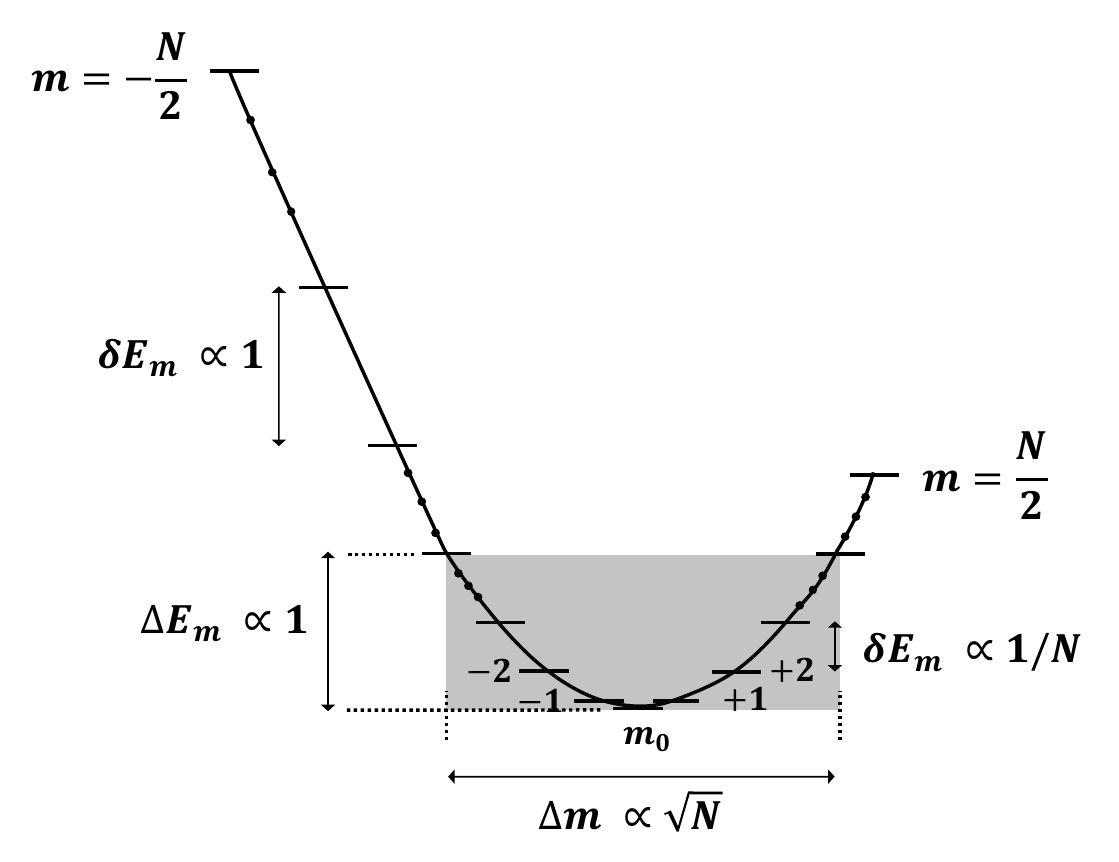}
	\caption{The energy levels for $\gamma > 1/2$ which are labeled by  quantum number $m$ and denoted by black bars.
	The ground state is at $m_0$. The energy gap $\delta E$ between neighboring levels around $m_0$ is $\propto 1/N$.
	All the energy levels marked by the gray area are quasi-ground states, whose energies become degenerate when $N\rightarrow\infty$.
	The width of the gray area is about $\sim \sqrt{N}$. The  energy gaps outside the gray area  remain finite even at the large $N$ limit. }
	\label{fig:2}
\end{figure}

When $\gamma>1/2$, the energy level is the lowest at $m_{0}$, which  is the largest integer smaller than or equal to $N/(4\gamma)$.
The highest energy level is at $m=-S$. Without loss of essential physics, we choose  $m_0=N/(4\gamma)$ by
choosing an appropriate value of $\gamma$.  The energy gap between the ground state and the first excited state is $2\gamma/N$;
the gap between the highest energy level and the second highest is $1+2\gamma-2\gamma/N$. The former approaches zero and
the latter remains finite at the limit $N\rightarrow \infty$. This means
that for $\gamma>1/2$ the energy gap $\delta E_m$ between two neighboring levels has different dependence on $N$ at different levels $m$.
We find that for a set of energy levels near the ground state the gap approaches zero and for others it remains finite at $N\rightarrow \infty$.
This is shown schematically in Fig.\;\ref{fig:2}. To accurately to describe this behavior, we define the energy gap $\Delta E_m$
between energy level $\ket{m}$ and the ground state $\ket{m_0}$ and we find that
\be
\Delta E_m=\frac{2\gamma }{N}(m-m_0)^2\,.
\ee
This means that any energy level $\ket{m}$ will approach the ground state energy in the thermodynamic limit $N\rightarrow \infty$
when $m$ satisfies
\begin{equation}\label{eq:7}
	|m-m_0|\sim O\left(N^{1/2-\delta}\right)\,,
\end{equation}
where $0<\delta<1/2$. These energy levels are located in the shadow area in Fig.\;\ref{fig:2}.
We call them quasi-ground states, which form a sub-Hilbert space $C$. At the limit $N\rightarrow \infty$, all these quasi-ground states
become the mean-field ground states on the black dashed line in Fig. \ref{fig:1}(b).

\begin{figure}[h]
	\centering
	\includegraphics[width=1\linewidth]{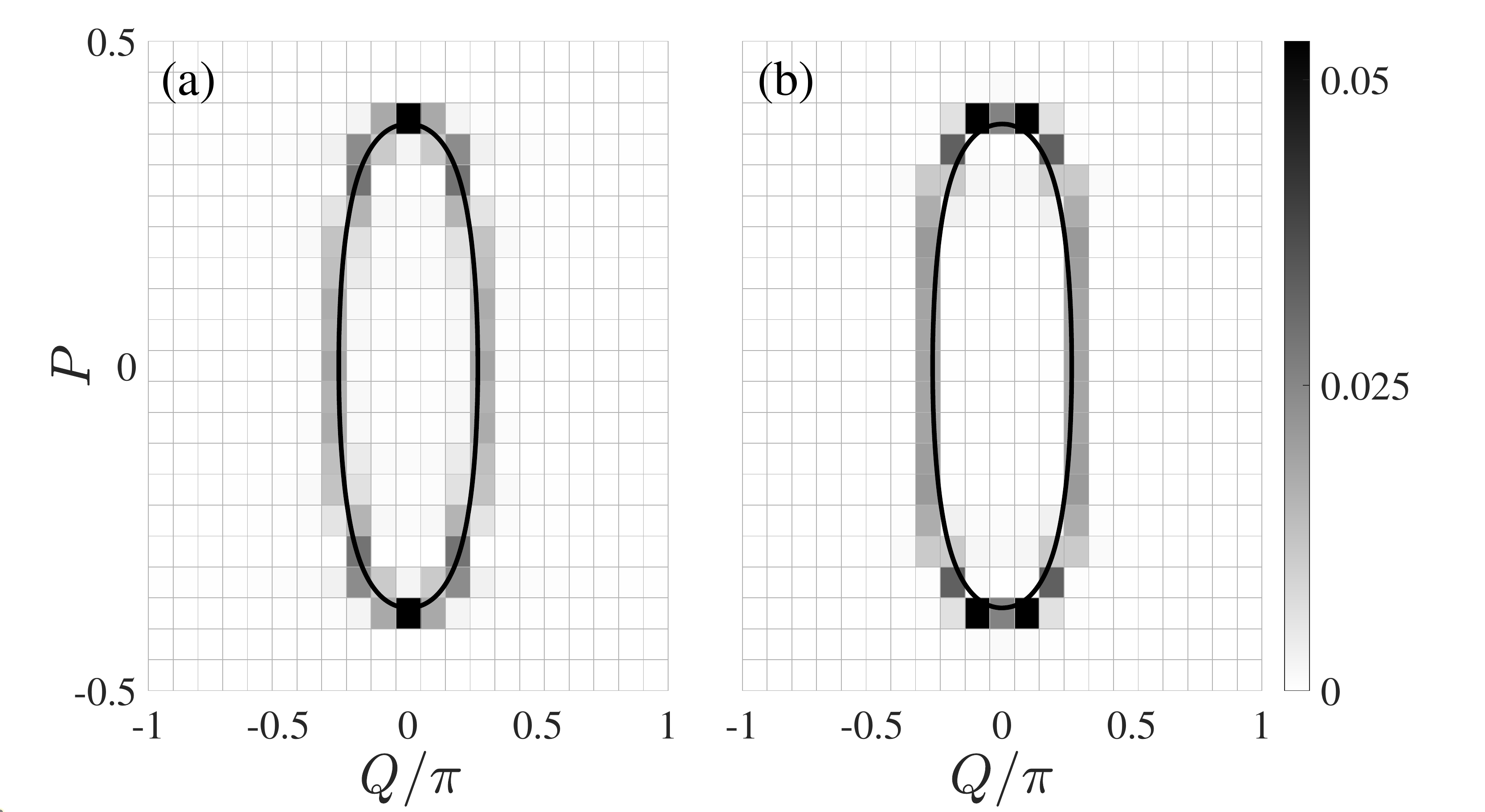}
	\caption{Energy eigenstates $\ket{m}$  in the quantum phase space for  $N=440$ and $\gamma=3/4$.
	(a) The true ground state; (b) the $29$th exited state.
	The black lines represent the mean-field infinite degenerate ground states.}
	\label{eng}
\end{figure}

\subsection{Eigenstates in quantum phase space}\label{sec:2C}
As discussed above,  in the case of $\gamma > 1/2$
there are many quasi-ground states in the quantum model $\hat{H}$, which correspond to the infinitely many ground states
in the mean-field model. Such a correspondence becomes more apparent and insightful when we plot these quasi-ground states
in the quantum phase space (see Fig. \ref{eng}).

The  quantum phase space is obtained by dividing the mean-field phase space in Fig. \ref{fig:1}
into Planck cells. The size of a Planck cell, which is denoted by $(Q_i,P_j)$, is $1/N$.
A Wannier function $\left\{\ket{Q_i,P_j}\right\}$, which is localized along both $P$ and $Q$ directions,
is assigned for each Planck cell and all the Wannier functions form a complete set of
orthonormal basis\cite{neumann1929beweis, han2015entropy, jiang2017quantum, fang2018quantum,Zhenduo2021Microscope}.
 An energy eigenstate $\ket{m}$ is then projected to the quantum phase space as
 \be
 \ket{m}=\sum_{i,j}\ket{Q_i,P_j}\braket{Q_i,P_j|m}\,.
 \ee
 What is plotted in Fig. \ref{eng} is $|\langle Q,P|m\rangle|^2$. Two examples are shown in Fig. \ref{eng}: one is
 the true ground state $\ket{m_0}$ and the other is a quasi-ground state.  It is clear from the figure that
 both eigenstates are concentrated around the mean-field ground states, which are marked by black solid lines in Fig. \ref{eng}.

\section{\label{sec:3} LOGARITHMIC TIME CRYSTAL}	
The quantum time crystal can be constructed as follows. We choose an eigenstate $\ket{m_1}$ with
\begin{equation}\label{eq:10}
	m_1=m_0-\Big\lfloor\sqrt{\frac{N}{\log N}}\Big\rfloor\,,
\end{equation}
where $\lfloor x \rfloor$ is the largest integer $\le x$. This state $\ket{m_1}$ certainly belongs to
the sub-Hilbert space $C$ as $\Delta E_{m_1}=2\gamma/\log N$ approaches zero when $N\rightarrow \infty$.
We construct a superposition state $\ket{\Phi_0}=(\ket{m_0}+\ket{m_1})/\sqrt{2}$. It will evolve with time as
 \begin{equation}\label{eq:11}
 	|\Phi(t)\rangle=\frac{e^{-iE_{m_0}t}}{\sqrt{2}}\left[
 	|m_0\rangle+\exp\left(-\frac{i2\pi t}{T_0\log{N}}\right)|m_1\rangle
 	\right]\,,
 \end{equation}
where $T_0=\pi/\gamma$ is a constant with order $O(1)$ and we have ignored the minor difference between $x$ and $\lfloor x \rfloor$.
It is clear that this quantum state will oscillate with a period of $T_0\log(N)$.

In real experiments, it is hard to prepare a state that is  a superposition of  only two eigenstates.
We consider a double-Gaussian superposition state $|\Psi(0)\rangle=\sum_{m}c_m|m\rangle$,  where
\be
\label{eq:14}
	c_{m} = \frac{1}{\sqrt{2\pi \sigma}}\Big[e^{-\frac{(m-m_{0})^{2}}{4\sigma^2}}
	+e^{-\frac{(m-m_{1})^{2}}{4\sigma^2}}\Big]\,,
\ee
where $\sigma$ is the width of the Gaussian distribution.  Since we do not want the Gaussian distributions to spread out the whole sub-Hilbert
space $C$ of quasi-ground states,  the width $\sigma$ should be much smaller than $|m_1-m_0|$.
For simplicity, we choose $\sigma\sim O(1)$.

\begin{figure}[h]
	\centering
	\includegraphics[width=0.9\linewidth]{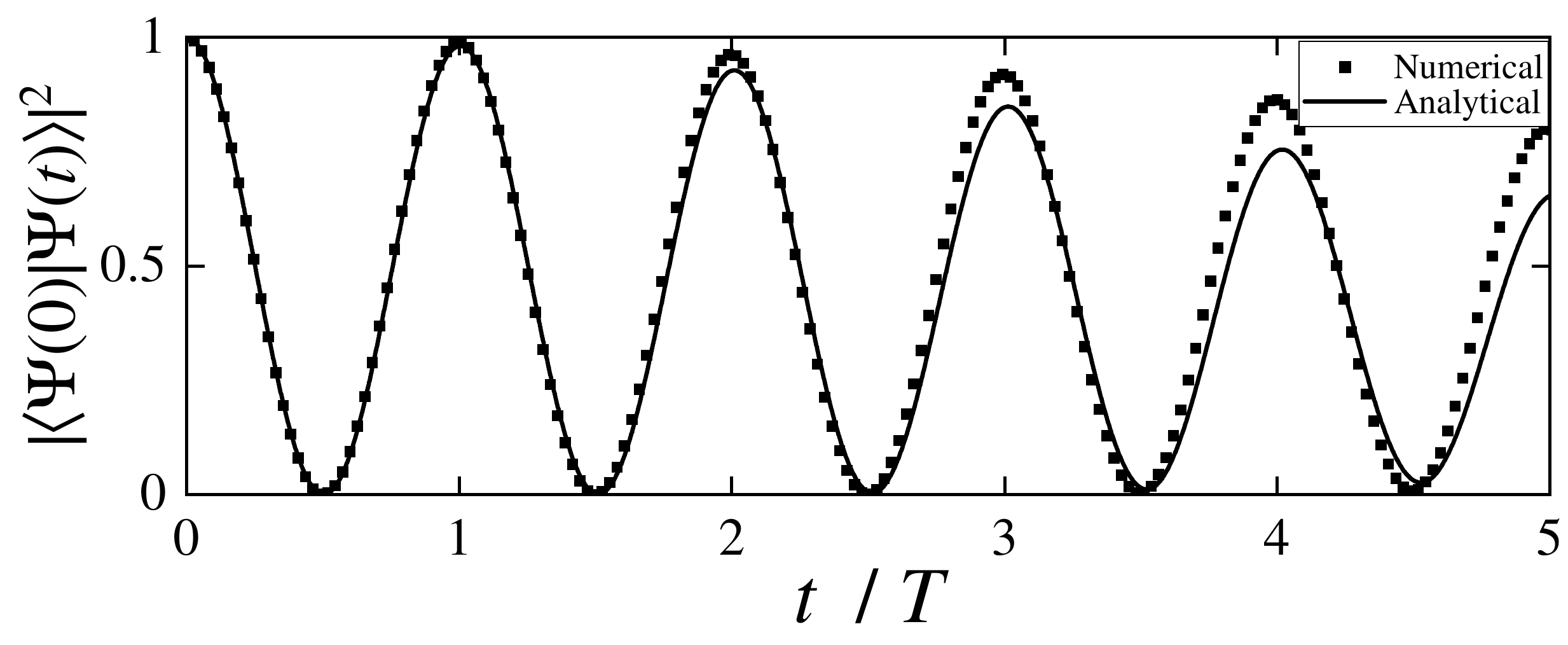}
	\caption{The oscillations of  logarithmic time crystal (\ref{eq:14}).  The solid line is the analytical result in Eq.(\ref{eq:16}) and
	the squares are numerical results obtained by directly summing Eq.\eqref{eq:15}.
	$N=29240$, $m_1=m_0+91$, $\sigma = 1$. }
	\label{fig:4}
\end{figure}

The overlap between $|\Psi(t)\rangle$ and $|\Psi(0)\rangle$ can be evaluated as
\begin{align}\label{eq:15}
	\langle \Psi(0) |\Psi(t)\rangle = \sum_{m=-N/2}^{N/2} |c_m|^2 \exp\left(-i\Delta E_m t\right)\,.
\end{align}
When $N$ is very large, we can replace the summation by integral,
\begin{align}\label{eq:16}
	&\langle \Psi(0) |\Psi(t)\rangle \approx \int_{-\infty}^{\infty}d m\; |c_m|^2 \exp\left(-i\Delta E_m t\right) \notag\\
		= & A(t)\left\{ 1 + \exp\left[-\frac{N^2}{4\Sigma^2(t)\log^2 N}\right]\exp\left(-i\frac{2\pi t}{T(t)}\right)\right\}\,,
\end{align}
where
\begin{align}\label{eq:17}
	T(t) &= T_0\log N \left(
	1+\frac{64\gamma^2\sigma^4 t^2}{N^2}
	\right),\notag\\
	\Sigma(t) &= \sqrt{\left(\frac{N}{8\gamma \sigma t}\right)^2+{ \sigma^2}},\notag\\
	A(t) &= \frac{1}{2}\sqrt{\frac{N}{{N}+8i{\gamma \sigma^2 t}}}.
\end{align}
It is clear from $T(t)$ that when  $t\ll O(N)$ the system oscillates with a period of $T_0\log N$.
$A(t)$ shows that the amplitude of the oscillation  will gradually decrease with time.
However, within the time interval $0<t\ll O(N)$ we should be able to observe many periods of oscillations since
the period of oscillation is of order $O(\log (N))$. In addition, the oscillations are enveloped by a Gaussian function
with width $\Sigma(t)$ which increases with time. Instead of doing integral, one can also do
the summation in Eq.\eqref{eq:15} numerically.  Both results are shown in Fig.\;\ref{fig:4} for comparison;
they agree very well.

\section{EXPERIMENTAL Scheme}
\label{sec:4}
\begin{figure}
	\centering
	\includegraphics[width=0.8\linewidth]{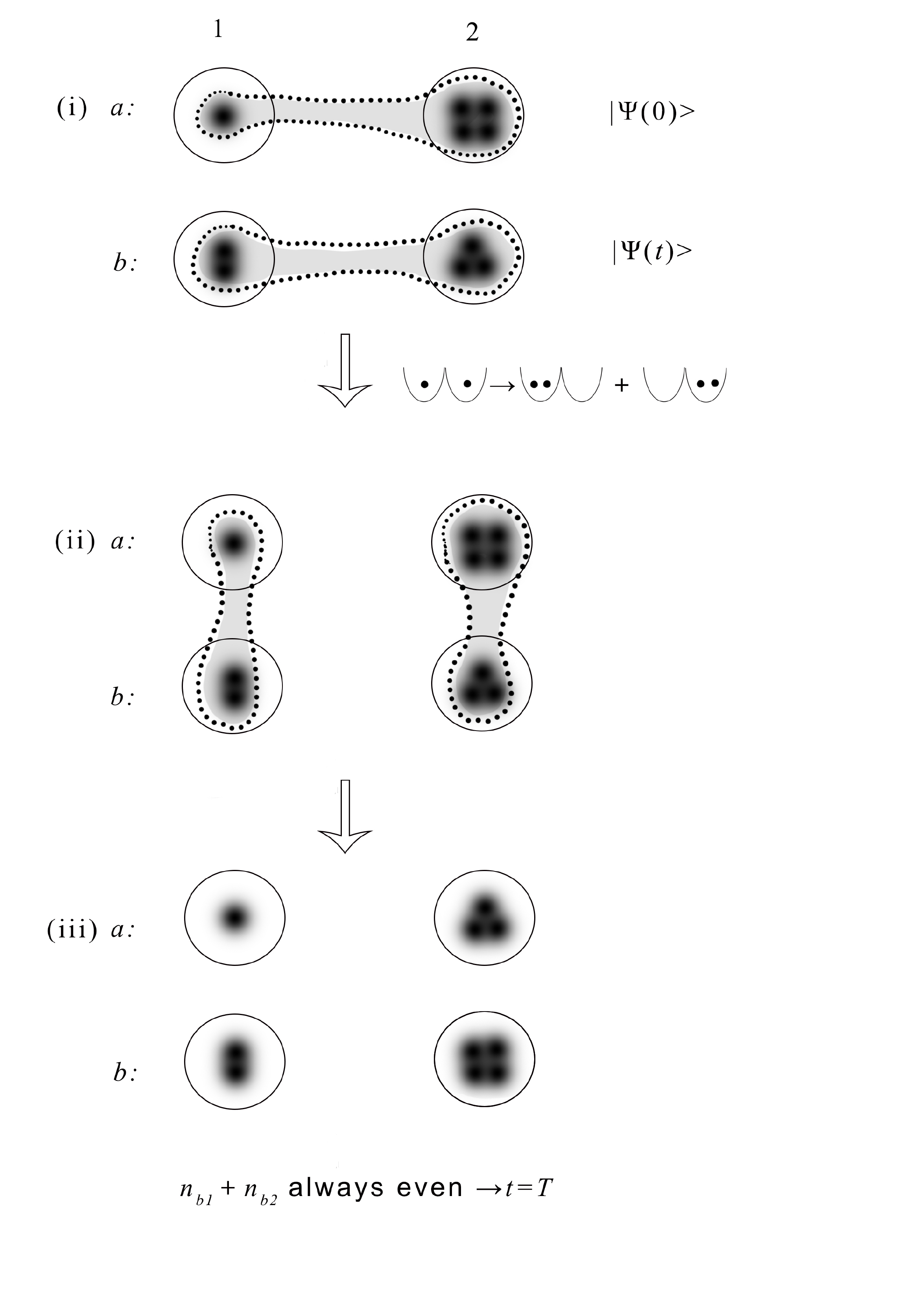}
	\caption{The experiment scheme. (i) Two identical boson systems $a$ and $b$; each has two modes, denoted by $1,2$.
	Initially,  the two systems are independent and they are prepared in states, $|\Psi(0)\rangle$ and $|\Psi(t)\rangle$, respectively.
	(ii) We allow the atom tunneling between the modes $a_i$ and $b_i$ ($i=1,2$) to realize Hong-Ou-Mandel interference.
	(iii) We count particle number parity in system $b$. Repeat the above process for many times.
	This scheme effectively measures $\left|\langle \Psi(0)| \Psi(t)\rangle\right|^2$. }
	\label{fig:experiment}
\end{figure}

In literature, the time crystal dynamics is usually illustrated with the correlation function of the polarization along the $z$ direction,
$\langle\hat{S}_z(0)\hat{S}_z(t)\rangle$ \cite{sacha2017time}.
For the quantum time crystal Eq.(\ref{eq:11}),  we have
\begin{align}\label{eq:12}
\langle\hat{S}_z(0)\hat{S}_z(t)\rangle
&= A_{0,+} e^{-i\delta E_{m_{0}+1}t}+A_{0,-} e^{-i\delta E_{m_{0}}t}\notag\\
&\quad +A_{1,+} e^{-i\delta E_{m_{1}+1}t}+A_{1,-} e^{-i\delta E_{m_{1}}t}\,,
\end{align}
where $A_{j,\pm}= (N/2+1)N/2-m_{j}(m_{j}\pm 1)$, $j=0,1$.
The four characteristic frequencies $\delta E_{m_{j}\pm 1}$ are just the energy gaps between
the neighboring levels at  the two states $|m_{0}\rangle$ and $|m_{1}\rangle$. Therefore, we have $\delta E_{m_{j}\pm 1}\ll O(1/\log N)$ and
the oscillation period is much longer than $\log N$.
So the correlation function Eq.(\ref{eq:12}) can not be used to observe the dynamic behavior of
our logarithmic quantum time crystal. We propose to use the experimental scheme shown in Fig.\;\ref{fig:experiment},
where $\left|\langle \Psi(0)| \Psi(t)\rangle\right|^2$ is measured.

In Fig.\;\ref{fig:experiment}, there are two identical two-mode boson systems, which are denoted as $a$ and $b$,
created with atoms and optical devices.
The tunneling between system $a$ and $b$ can be tuned: initially the optical potential barrier between them is high
so that there is no tunneling; the barrier is lowered in the later stage of the experiment to allow Hong-Ou-Mandel interference.
Initially when the tunneling is off, both system $a$ and $b$ have the same number of bosons and
evolve with time under the Hamiltonian Eq.\;\ref{eq:1} .

A swap operator $\hat{V}$ between these two systems is defined as
 \be
 \hat{V}|\psi\rangle_a\otimes |\phi\rangle_b = |\phi\rangle_a\otimes |\psi\rangle_b\,,
\ee
where $|\cdot\rangle_{a}$ and  $|\cdot\rangle_{b}$ are states of systems $a$ and $b$.
It is easy to see that $\langle \hat{V} \rangle = \left|\langle \psi| \phi\rangle\right|^2$, i.e., the expectation value of the
swap operator $\hat{V}$ is the overlap between two quantum states $\ket{\psi}$ and $\ket{\phi}$.
As $\hat{V}$ has only two eigenvalues $\pm 1$,  measuring the expectation of operator $\hat{V}$ requires
statistically counting its eigenvalues within a large ensemble.
 In our system,  $\hat{V}= \hat{V}_1\otimes\hat{V}_2$, where $\hat{V}_{i}$ is the swapping on
 individual modes $i=1,2$. So measuring  $\hat{V}$ can be achieved by measuring $\hat{V}_1$ and $\hat{V}_2$.

The symmetric eigenstate $|S\rangle$ and asymmetric eigenstate $|AS\rangle$ of $\hat{V}_{i}$
can be expressed in terms of  Fock states as follows
\begin{align}\label{eq:21}
	|S\rangle &= \text{superpose} \left\{(\hat{a}_i^\dagger-\hat{b}_i^\dagger)^{2\alpha}(\hat{a}_i^\dagger+\hat{b}_i^\dagger)^{\beta}|\text{vac}\rangle \right\},\\
	\label{eq:211}
	|AS\rangle &= \text{superpose} \left\{(\hat{a}_i^\dagger-\hat{b}_i^\dagger)^{2\alpha-1}(\hat{a}_i^\dagger+\hat{b}_i^\dagger)^{\beta}|\text{vac}\rangle \right\}\,,
\end{align}
where $\hat{a}_i^\dagger (\hat{b}_i^\dagger)$ refers to the creation operator on mode $a_i (b_i)$ and  $\alpha$ and $\beta$ are positive integers.
We perform the following transformation,
\be
\left(	\hat{a}_i^\dagger + \hat{b}_i^\dagger\right)/\sqrt{2}\rightarrow \hat{a}_i^\dagger\,,~~~\left(	\hat{a}_i^\dagger - \hat{b}_i^\dagger\right)/\sqrt{2}\rightarrow \hat{b}_i^\dagger\,.
\ee
This transformation can be realized experimentally with atoms tunneling between $a_i$ and $b_i$
under a Bose-Hubbard Hamiltonian \cite{islam2015measuring}.  This is essentially  Hong-Ou-Mandel interference.
After the transformation, the eigenstates of $\hat{V}_{i}$ become
\begin{align}\label{eq:22}
	|S'\rangle &= \text{superpose} \left\{(\hat{b}_i^\dagger)^{2\alpha}(\hat{a}_i^\dagger)^{\beta}|\text{vac}\rangle \right\},\notag\\
	|AS'\rangle &= \text{superpose} \left\{(\hat{b}_i^\dagger)^{2\alpha-1}(\hat{a}_i^\dagger)^{\beta}|\text{vac}\rangle \right\}.
\end{align}
This means that  observing even(odd) particle numbers in mode $b_i$ corresponds to eigenvalues $+1(-1)$ of $\hat{V}_{i}$.
Hence if the particle numbers in mode $b_1$ and $b_2$ have the same parity, we
know the measured quantity of $\hat{V}$ is $+1$; otherwise, it is $-1$.

The basic experimental steps are shown in Fig.\;\ref{fig:experiment}. Initially, we prepare system $b$ in state $|\Psi(0)\rangle$ and
let it evolve for a period of time $t$ so the state becomes $|\Psi(t)\rangle$. At this same moment,
we prepare system $a$ in state $|\Psi(0)\rangle$. Then we turn off the tunneling between different modes and turn on the
the tunneling between system $a$ and $b$, realizing Hong-Ou-Mandel interference. Finally, we observe the number of particles
in both modes of system $b$. By repeat this experiment many times for different $t$,
 $\langle \hat{V} \rangle = \left|\langle \Psi(0)| \Psi(t)\rangle\right|^2$ is obtained.

\section{\label{sec:level4} Discussion And Conclusion}
Before concluding, we need to mention that performing perturbed external fields  \cite{huang2018symmetry} actually fails to produce a logarithmic time crystal state. Because the field, like $g\hat{S}_{z}$, can only superpose the state $|m_{0}\rangle$ with its nearest levels. To be honest, how to realize logarithmic time crystal is still an open question for us. But we do have some speculations. 
In solid physics, electrons can be excited from the band bottom to the band top by the phonons with specific momentum and energy, where the momentum of phonons is tuned by the structure of Brillouin-zone. In our model, the system can be excited by a laser from $|m_0\rangle$ to $|m_1\rangle$ when the frequency of laser is equal to $\Delta E_{m_1} = E_{m_1}-E_{m_0}$. But only controlling the frequency cannot assure that the intermediate states are not superposed. So, we need to speculate that there may be some extra characteristics (like Brillouin-zone with respect to electrons) about our logarithmic time crystal state to rule the momentum of laser so that it can be achieved. 

In conclusion, we have studied a two-mode boson system of $N$ particles. In a certain parameter regime, this system has many quasi-ground states, which forms a sub-Hilbert space $C$ and become degenerate with the true ground state in the thermodynamic limit $N\rightarrow\infty$. We have been able to  superpose these quasi-ground states and construct a quantum time crystal that oscillates with a period that is logarithmically
proportional to the number of particles, $T\sim O(\log(N))$.  Such a logarithmic dependence makes the period short enough for possible experimental observation even when the number of particle is large. A experimental scheme based on Hong-Ou-Mandel interference has been proposed.

\begin{acknowledgments}
We thank Yijie Wang for helpful discussion. This work is supported by the National Key R\&D Program of China (Grants No.~2017YFA0303302, No.~2018YFA0305602), National Natural Science Foundation of China (Grant No. 11921005), and
Shanghai Municipal Science and Technology Major Project (Grant No.2019SHZDZX01).
\end{acknowledgments}



\nocite{*}

%

\end{document}